\documentclass[journal]{IEEEtran}
\newcommand{\RNum}[1]{\uppercase\expandafter{\romannumeral #1\relax}}
\usepackage{mathrsfs}
\usepackage{amssymb}
\usepackage[mathcal]{euscript}
\usepackage{diagbox}
\usepackage{cite}
\usepackage[T1]{fontenc}
\usepackage{graphicx}
\usepackage{CJKutf8}
\usepackage{makecell}
\usepackage{psfrag}
\usepackage{url}
\usepackage{stfloats}
\usepackage{amsmath}
\usepackage{array}
\usepackage{float}
\usepackage{multirow} 
\usepackage{hyperref}
\usepackage{amsthm}
\usepackage{color}
\usepackage{multicol}
\usepackage{stfloats}
\usepackage{enumerate}
\usepackage{subfigure}
\usepackage{booktabs}  

\allowdisplaybreaks[4]
\usepackage[ruled]{algorithm2e}
\usepackage{algorithmic} 
\usepackage{xurl}
\hypersetup{breaklinks=true}

\ifCLASSINFOpdf
\else
\fi
\begin{document}
\sloppy
\begin{CJK}{UTF8}{gbsn}
\title{When Graph Meets Retrieval Augmented Generation for Wireless Networks: A Tutorial and Case Study}

\author{Yang Xiong, Ruichen Zhang, Yinqiu Liu, Dusit Niyato, \emph{Fellow, IEEE,}\\ Zehui Xiong, Ying-Chang Liang, \emph{Fellow, IEEE,} Shiwen Mao, \emph{Fellow, IEEE}
\thanks{Y. Xiong is with the Engineering Systems and Design Pillar, Singapore University of Technology and Design (e-mail: yang\_xiong@mymail.sutd.edu.sg)}
\thanks{R. Zhang, Y. Liu, and D. Niyato are with the College of Computing and Data Science, Nanyang Technological University, Singapore (e-mails:
ruichen.zhang@ntu.edu.sg; yinqiu001@e.ntu.edu.sg; dniyato@ntu.edu.sg).}
\thanks{Z. Xiong is with the Computer Science and Design Pillar, University of Technology and Design (e-mail: zehui\_xiong@sutd.edu.sg).}
\thanks{Y.-C. Liang is with the Center for Intelligent Networking and Communications (CINC), University of Electronic Science and Technology of China (UESTC), Chengdu 611731, China. (e-mail: liangyc@ieee.org).}
\thanks{S. Mao is with the Department of Electrical and Computer Engineering, Auburn University, Auburn, AL 36849, USA (e-mail: smao@ieee.org).}
}

\maketitle

\begin{abstract}
The rapid development of next-generation networking technologies underscores their transformative role in revolutionizing modern communication systems, enabling faster, more reliable, and highly interconnected solutions. However, such development has also brought challenges to network optimizations. Thanks to the emergence of Large Language Models (LLMs) in recent years, tools including Retrieval Augmented Generation (RAG) have been developed and applied in various fields including networking, and have shown their effectiveness. Taking one step further, the integration of knowledge graphs into RAG frameworks further enhanced the performance of RAG in networking applications such as Intent-Driven Networks (IDNs) and spectrum knowledge maps by providing more contextually relevant responses through more accurate retrieval of related network information. This paper introduces the RAG framework that integrates knowledge graphs in its database and explores such framework's application in networking. We begin by exploring RAG's applications in networking and the limitations of conventional RAG and present the advantages that knowledge graphs' structured knowledge representation brings to the retrieval and generation processes. Next, we propose a detailed GraphRAG-based framework for networking, including a step-by-step tutorial on its construction. Our evaluation through a case study on channel gain prediction demonstrates GraphRAG’s enhanced capability in generating accurate, contextually rich responses, surpassing traditional RAG models. Finally, we discuss key future directions for applying knowledge-graphs-empowered RAG frameworks in networking, including robust updates, mitigation of hallucination, and enhanced security measures for networking applications.
\end{abstract}


\IEEEpeerreviewmaketitle

\section{Introduction}In recent years, Artificial Intelligence (AI)—especially generative Large Language Models (LLMs)—has progressed rapidly, revolutionizing various fields, including networking. Advanced LLMs, such as GPT\footnote{{https://openai.com/index/gpt-4/}}, Llama\footnote{{https://ai.meta.com/blog/meta-llama-3-1/}}, and Gemini\footnote{{https://deepmind.google/technologies/gemini/}}, have greatly expanded the potential of Artificial Intelligence-Generated Content (AIGC) across diverse applications. In industry, for instance, Singtel\footnote{{https://www.singtel.com/about-us/media-centre/news-releases/global-telco-ai-alliance-founding-parties-sign-agreement}}, SK Telecom\footnote{{https://www.sktelecom.com/en/press/press\_detail.do?idx=1612}}, and several companies established the Global Telco AI Alliance to co-develop and launch a multilingual Telco LLM, aiming to improve customer interactions through digital assistants and innovative AI solutions. In research, for example, \cite{demllm} proposed a framework for the democratized deployment of LLM agents on mobile devices; and \cite{idnllm} demonstrated how Intent-Driven Networks (IDNs) can leverage LLMs and RAG to automate network configurations, translating user intents into CLI (Command-Line Interface) commands while ensuring accuracy through a Network Digital Twin (NDT).

\begin{table*}[ht]
    \centering
    \caption{Review of existing applications of LLMs}
    \begin{tabular}{c} 
        \includegraphics[width=0.9\textwidth]{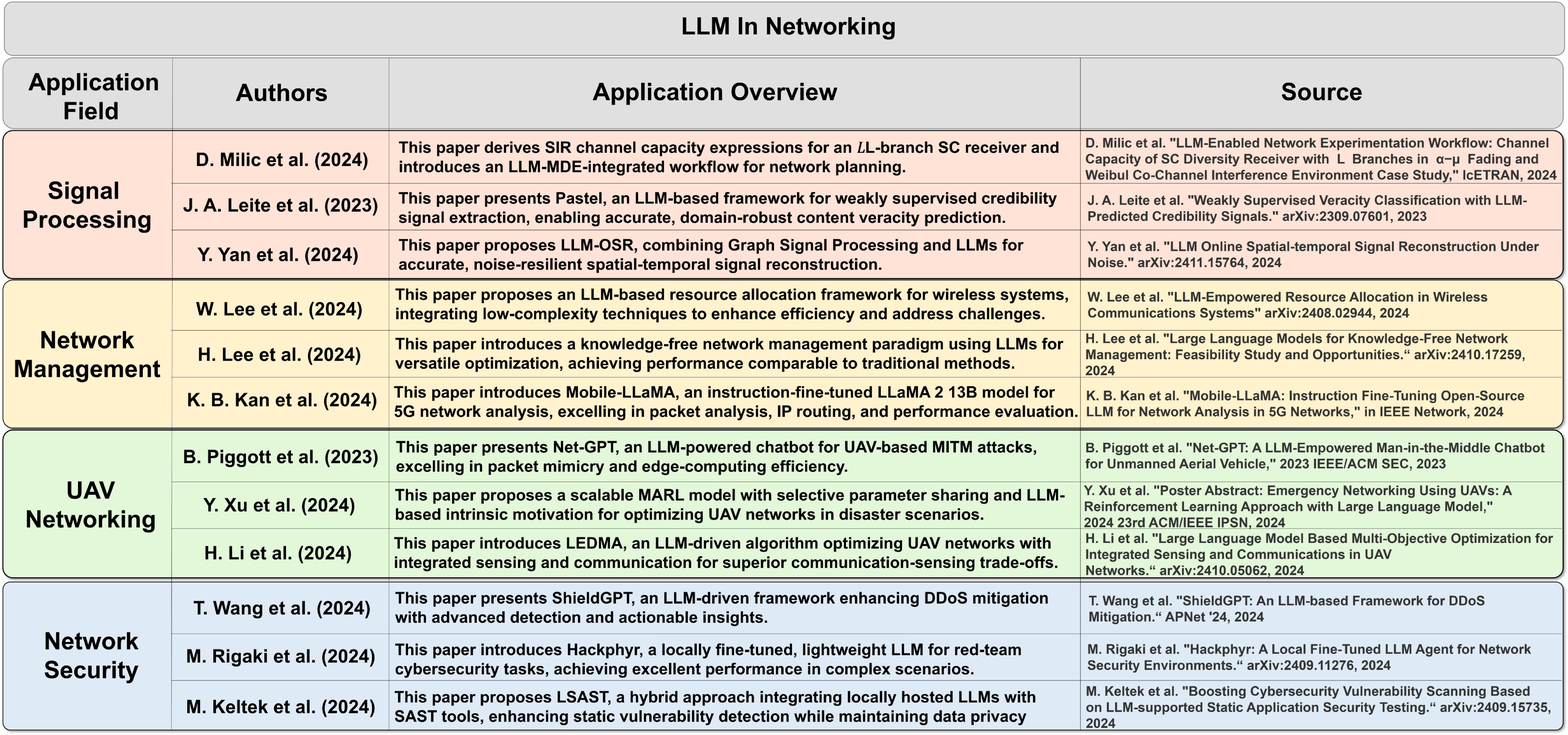} 
    \end{tabular}
    
    \label{tab:1}
\end{table*}
Despite these advancements, early LLMs encountered notable limitations, particularly in specialized fields including networking, where real-time, domain-specific knowledge is essential. A major challenge was the LLMs’ restricted access to current, relevant data, limiting their ability to address complex, dynamic networking demands. To mitigate this, Retrieval-Augmented Generation (RAG) \cite{rag} has emerged as a solution. In RAG, an LLM is augmented by an external knowledge base that retrieves and supplies relevant information to guide the model’s response generation. Such characteristics of RAG facilitate the applications of LLMs in networking. For example, the authors in \cite{c} proposed an RAG-involved framework for UAV networking, enabling automated design of network structures and loss functions for spectrum estimation. Experimental results show that the framework achieves superior spectrum mapping accuracy, outperforming traditional methods like LSTM, while optimizing resource allocation and enhancing adaptability in dynamic environments. 

However, RAG faces its own challenges. In the networking context, RAG often struggles with obtaining high retrieval accuracy, processing complex queries, and handling large databases, especially when intricate relationships between data points are needed. One of the challenges lies in accurately interpreting relationships within raw network data, particularly when critical information is embedded mid-dataset. For example, in a dataset describing connections among devices, RAG performs better when retrieving information about the devices that appear at the beginning or the end of the dataset. But for the devices that appeared in the mid-dataset, RAG might ignore them or retrieve inaccurate information. This limitation can lead to incomplete or inaccurate network mappings. To address these gaps, knowledge graphs were introduced into the construction of RAG frameworks. Different from standard RAG, which relies on retrieving flat text chunks, constructing a structured knowledge graph that organizes data as entities and relationships allows the system to capture complex contextual relationships, reduce the hallucination effect in LLMs, and provide targeted retrievals that align better with network-specific challenges. 

Building on these developments, this paper provides a forward-looking perspective on applying RAG frameworks with knowledge graphs to next-generation networking. We investigate how the knowledge-graphs-empowered RAG framework's structured knowledge representation and generation capability can address specific challenges in networking. In our case, we leverage the natural advantage of graph structures in representing the connection between devices to retrieve the channel information between transmitters and receivers in the network for channel gain prediction. To guide our analysis, we focus on the following questions:

\textbf{Q1:} Why is the knowledge-graphs-empowered RAG framework particularly well-suited to addressing networking challenges?

\textbf{Q2:} What specific networking issues can the knowledge-graphs-empowered RAG framework help solve?

\textbf{Q3:} How can the knowledge-graphs-empowered RAG framework be effectively applied to tackle these issues?

To answer these questions, this paper examines the structures and functionalities of vanilla RAG and knowledge-graphs-empowered RAG framework, and their applications in networking. We propose a GraphRAG-based framework and validate its effectiveness through a detailed case study. \emph{To the best of our knowledge, this is the first article to explore the application of the knowledge-graphs-empowered RAG in networking.} Our contributions are summarized as follows:
\begin{itemize}

\item We first review the structure and existing applications of RAG in networking to establish an understanding of RAG's applications in networking. Then we discuss its limitations and introduce the advantages of knowledge-graphs-empowered RAG over vanilla RAG.

\item We present the GraphRAG framework with advanced retrieval and generation capabilities that provide the framework with advantages in networking applications such as extracting the interconnections among devices in the spectrum map. We also present a step-by-step guide for its construction and application within networking contexts.

\item We evaluate the effectiveness of the proposed GraphRAG framework through our case study that applies the GraphRAG framework to predict the channel gain between the transmitters and receivers given their locations and the Channel Knowledge Map (CKM). We first provide a basic validation of the advantages of GraphRAG over vanilla RAG, then validate the effectiveness of the GraphRAG framework through the channel gain prediction experiment.

\end{itemize}

\section{Overview of Retrieval Augmented Generation and GraphRAG in Networking}

\begin{figure*}[!ht]       
	\centering
        
        \includegraphics[width=0.9\textwidth]{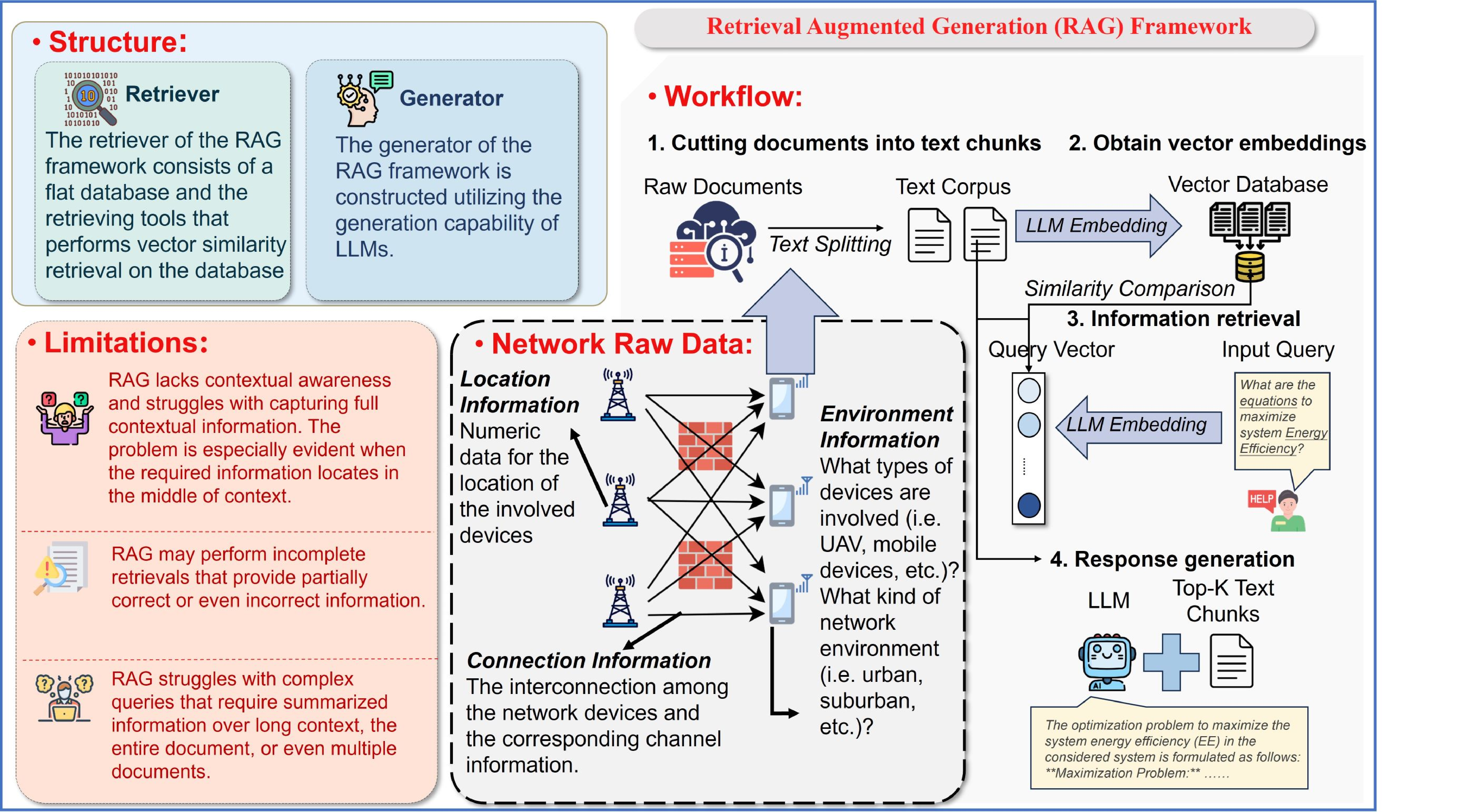}
        
	\caption{Structure, applications, and limitations of baseline RAG. Despite showing its effectiveness in solving networking problems, baseline RAG still has its limitations and can be further improved.}   
	\label{FIG:1}
\end{figure*}

\subsection{Retrieval Augmented Generation in Networking}
To enable efficient networking as it evolves toward the next generation, it is crucial to address its inherent challenges where traditional methods may face limitations, such as scalability, sustainability, and reliability \cite{6g}. To overcome these limitations, more powerful and adaptable tools are required.

\subsubsection{Large Language Models in Networking}
LLMs have emerged as powerful tools in networking, offering innovative solutions for automation, optimization, and intelligent decision-making across various domains. As we illustrate in Table 1, studies have been conducted to explore the applications of LLMs in networking across various network functions. However, despite their impressive capabilities, LLMs are not without limitations\footnote{{https://www.enterprisedb.com/blog/limitations-llm-or-why-are-we-doing-rag}}. They often struggle with accessing and utilizing up-to-date, domain-specific information, as their knowledge is confined to the data available during their training. This limitation poses challenges in dynamic and specialized fields like networking, where real-time data and context-specific understanding are critical.

\subsubsection{RAG Structure}
To address these gaps, Retrieval-Augmented Generation (RAG) has been introduced as a complementary approach. As illustrated in Figure 1, the RAG framework, powered by advances in LLMs, offers a promising solution, facilitating intelligent, real-time knowledge retrieval and dynamic optimization across the network. RAG combines pre-trained retrieval models with generative LLMs to enhance the relevance and contextual accuracy of responses. The system functions by retrieving pertinent information from a large database and generating a response informed by both the query and the retrieved content. RAG has two core components:

\begin{itemize}
\item \textbf{Retriever}: The retriever locates the most relevant information within the database, using a neural language model to embed both the query and documents as vectors. Based on vector similarity, it ranks and retrieves top text chunks. Due to LLMs' context length limitations, documents are split into manageable chunks and stored in a flat-structured database.

\item\textbf{Generator}: The generator uses the retrieved text chunks and input query to produce coherent, contextually relevant responses. This generative LLM component delivers precise answers, summaries, or other outputs that align with the specific needs of the networking task at hand.
\end{itemize}

\subsubsection{RAG Applications in Networking}
The structure of RAG has enabled RAG systems with several strengths and their applications to address challenges encountered in networking:

\begin{itemize}
\item\textbf{Intelligent AI Collaboration}: RAG has been applied to assist effective resource allocation through intelligent collaborations with multiple AI agents. For example, the authors in \cite{satelite} proposed a framework that incorporates an RAG-based generative AI agent with MoE-based PPO for satellite communication network optimization. The RAG agent formulates optimization models interactively, achieving accurate results with a cost of 2500 tokens per retrieval. MoE-PPO models process these formulations, minimizing human error and improving decision-making accuracy.

\item\textbf{Energy Consumption Optimization}: RAG has also been introduced in providing energy-efficient solutions in networking. Utilizing their strong analyzing and generating capabilities, RAG can help provide an overview of the optimization problems. For example, the framework proposed in \cite{lowcarbon} combines RAG-enabled LLMs and Generative Diffusion Models (GDMs) to address carbon emission optimization. RAG retrieves relevant academic data on carbon reduction, supporting LLM decision-making and guiding GDMs in generating strategies. Experimental results show a 17.97\% performance improvement over the DRL-PPO framework. 

\item\textbf{Enhanced Reliability}: RAG has also shown its effectiveness in enhancing the reliability of networks. By enabling real-time, context-aware decision-making, RAG helps provide reliable information about the network environment. For example, the authors in \cite{mobileai} introduced a 6G AI agent based on the LangChain framework, incorporating RAG and GPT-4 Turbo. Its performance is assessed across three autonomous driving scenarios, demonstrating effectiveness in real-time road updates, network configuration for tailored requests, and D2D vehicle communication. Results confirm its ability to enhance network reliability and performance significantly. 
\end{itemize}

\subsubsection{RAG Limitations}
Despite RAG’s advancements in enhancing generative model performance through retrieval, it has certain limitations:

\begin{itemize}
\item\textbf{Lack of Contextual Awareness}: RAG can struggle with capturing full contextual information, especially with interrelated entities across multiple document sections or crucial data embedded in the middle of text\footnote{{https://www.databricks.com/blog/long-context-rag-performance-llms}}. Experiments in \cite{ragcontext} show that when using GPT-3.5-Turbo\footnote{{https://platform.openai.com/docs/models/gpt-3-5-turbo}} for multi-document QA tasks, performance drops by over 20\% if key information is positioned mid-context rather than at the beginning or end. This decline is more pronounced as the number of retrieved documents increases, highlighting RAG's limited contextual awareness. In the context of networking, this weakness in missing information located in the middle of the database can raise problems with inaccurate mapping of network information, such as the connection information among devices.

\item\textbf{Incomplete Retrievals}: RAG may retrieve irrelevant or partially relevant data due to its lack of structured understanding of relationships between data points\footnote{{https://writer.com/blog/vector-based-retrieval-limitations-rag/}}. As experiments in \cite{ragacc} demonstrate, irrelevant content within the same topic can significantly impact LLM performance, decreasing generation accuracy by over 10\%. This issue underscores RAG's difficulty in consistently providing comprehensive and accurate outputs. In the context of network optimization, such inaccurate retrieval can provide significantly negative influences on the decision-making process.

\item\textbf{Challenges with Complex Queries}: RAG has limitations when handling complex queries that require synthesizing information from multiple sources or summarizing large documents\footnote{{https://microsoft.github.io/graphrag/\#graphrag-vs-baseline-rag}}. For instance, multi-document summarization experiments in \cite{ragsummary} show that models including T5\footnote{{https://huggingface.co/docs/transformers/model\_doc/t5}}, BART\footnote{{https://huggingface.co/docs/transformers/model\_doc/bart}}, and PEGASUS\footnote{{https://huggingface.co/docs/transformers/model\_doc/pegasus}} achieve low ROUGE-L\footnote{{https://klu.ai/glossary/rouge-score}} scores (below 0.2) on complex summarization tasks, indicating difficulty in generating cohesive summaries across related content. In the context of networking, lacking the capability to handle complex queries and summarize information can result in difficulties in obtaining a comprehensive overview of the network environment.
\end{itemize}

\begin{figure*}[!ht]       
	\centering
        \includegraphics[width=0.9\textwidth]{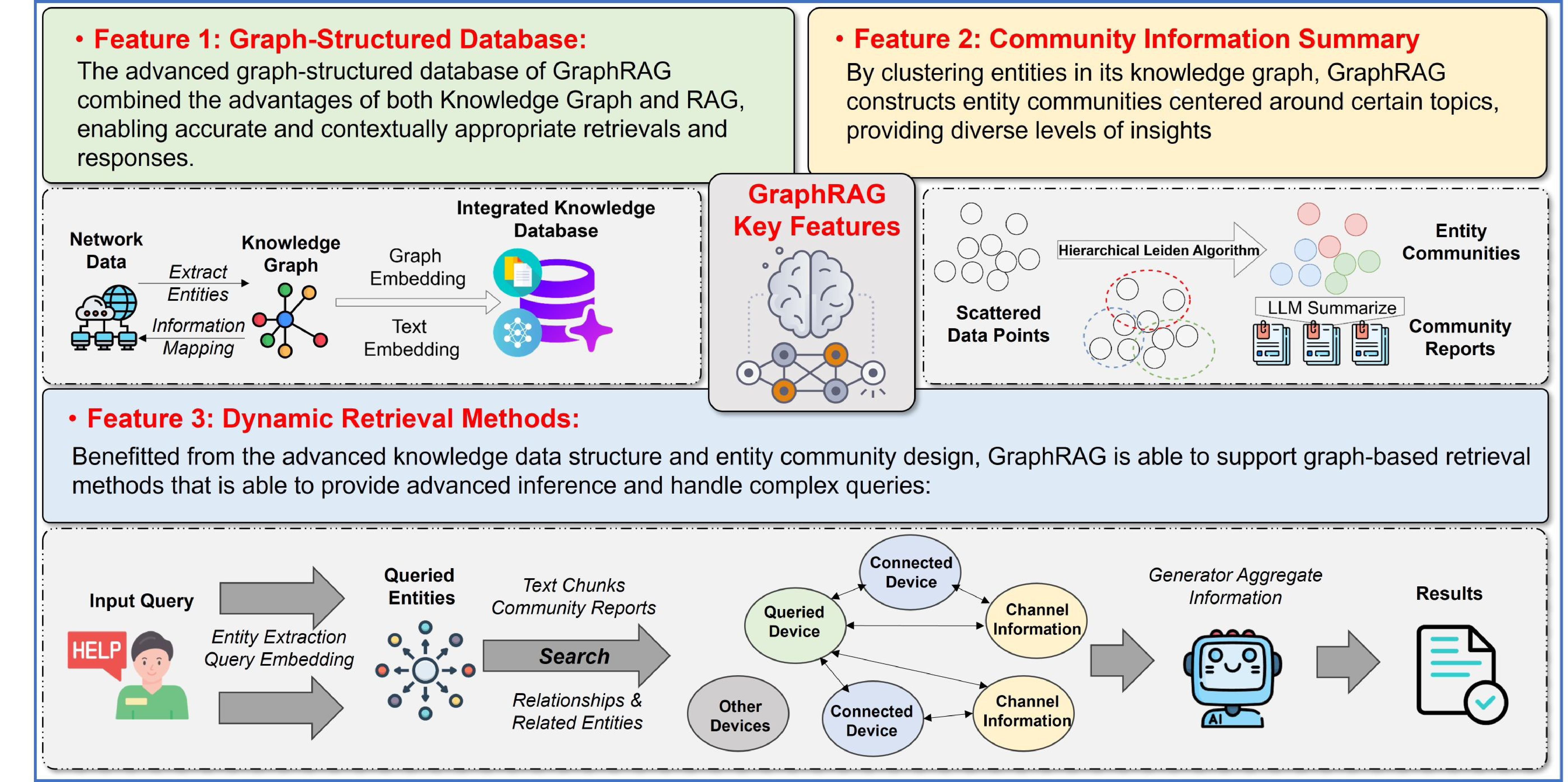}
	\caption{Key modifications in GraphRAG in comparison with the vanilla RAG. Integrating the structured knowledge graph into its database, GraphRAG takes one step further from vanilla RAG and is equipped with several advantages.}   
	\label{FIG:2}
\end{figure*}

\subsection{GraphRAG: Take One Step Further}
Despite previous enhancements, RAG's effectiveness in query-focused abstractive summarization over entire corpora remains limited. To address this, GraphRAG combines the strengths of RAG and knowledge graphs, overcoming each framework's limitations while enhancing its advantages.

\subsubsection{Database Structure}
 GraphRAG builds upon RAG's core features, adding a graph-structured database where entities and relationships are extracted by LLMs.\footnote{{https://www.microsoft.com/en-us/research/blog/graphrag-new-tool-for-complex-data-discovery-now-on-github/}} Flat text chunks are mapped to entities and relationships within the knowledge graph, with vector embeddings facilitating retrieval. Furthermore, entities are organized into topic-centered communities through clustering algorithms, enabling retrieval from varying levels of abstraction and providing multi-angled insights into the database. In network applications, for example, such database structure can be obtained through the extraction of the interconnections among network devices, with entities representing the devices and relationships (edges) representing the connections.

\subsubsection{Enhanced Retrieval Methods}
 GraphRAG supports two synergistic retrieval modes: Local Search and Global Search\footnote{{https://microsoft.github.io/graphrag/query/overview/}}. For Local Search, this method retrieves entity-specific information by traversing the graph's nodes (entities) and edges (relationships), capturing complex interdependencies to respond effectively to specific queries. For Global Search, Community-based retrieval uses summaries of topic-centered communities to answer broader queries. For example, considering a knowledge graph representing the network environment, the retrieval starts from the queried devices in the knowledge graph, then moves along their connections, and finally incorporates the explored information to identify the internet information (e.g., channel information, interconnections) around the queried devices.

The key characteristics of GraphRAG can be summarized in Figure 2. As a validation of GraphRAG's advantages, experiments in \cite{graphrag} comparing GraphRAG and vanilla RAG on various generation tasks demonstrated that GraphRAG improves answer comprehensiveness, diversity, and empowerment by approximately 30\%, producing more contextually helpful responses. In networking optimization, GraphRAG offers distinct advantages, including:

\begin{itemize}
{\item\textbf{Enhanced Contextual Understanding:}}
GraphRAG leverages the inherent relationships between entities in a knowledge graph\footnote{{https://microsoft.github.io/graphrag/index/default\_dataflow/\#phase-2-graph-extraction}}, enabling a deeper and more accurate understanding of context. It is able to assess the relevance of retrieved documents based on their relationships and connections, leading to more contextually appropriate retrieval. This characteristic of GraphRAG is particularly helpful in networking as it enables the system to model and understand complex interdependencies between devices, configurations, and protocols, leading to more accurate and context-aware decision-making. 

{\item\textbf{Advanced Querying Capabilities:}}
GraphRAG supports advanced semantic querying\footnote{{https://microsoft.github.io/graphrag/query/overview/}}, allowing it to interpret and answer complex queries more effectively by understanding the relationships and hierarchies within the graph. GraphRAG is able to reason about the data and retrieve information that may not be explicitly stated but can be inferred from the known relationships. Furthermore, by utilizing the relational information embedded in the graph, GraphRAG can disambiguate queries and better handle complex requests. This helps in networking by enabling precise analysis of complex queries across diverse network data for better troubleshooting and optimization.

{\item\textbf{Comprehensive Insights:}}
By integrating diverse data sources and relationships, GraphRAG provides a holistic view of the targeted problem. The community structure\footnote{{https://microsoft.github.io/graphrag/index/default\_dataflow/}} and the dynamic knowledge integration capability further allows GraphRAG to hold a multi-level understanding of the documents and handle distantly connected entities or concepts flexibly. This aids networking by offering a holistic view of network systems, enabling flexible analysis of complex, interrelated components for improved decision-making and problem resolution.
\end{itemize}

\subsection{Lesson Learned}
Our exploration of the applications of LLMs, RAG frameworks, and GraphRAG frameworks reveals several key insights. LLMs emerged as powerful tools in assisting network automation, optimization, and intelligent decision-making. However, it lacks the capability to adapt to application scenarios where frequently updated knowledge is required. RAG complements LLM by processing unstructured data, integrating new knowledge, and generating responses based on the given context. However, RAG still struggles with capturing complex relationships and ensuring comprehensive retrieval, highlighting its limitations. GraphRAG integrates knowledge graphs into its database structure, obtaining advanced query and generation capabilities, effectively addressing the limitations, and outperforming vanilla RAG in generation tasks for over 30\%. By offering improved querying, improved inference, and a more robust, scalable solution for the optimization of next-generation wireless networks, GraphRAG shows its potential as a powerful tool for modern communication systems.

\begin{figure*}[!ht]       
	\centering
        \includegraphics[width=0.95\textwidth]{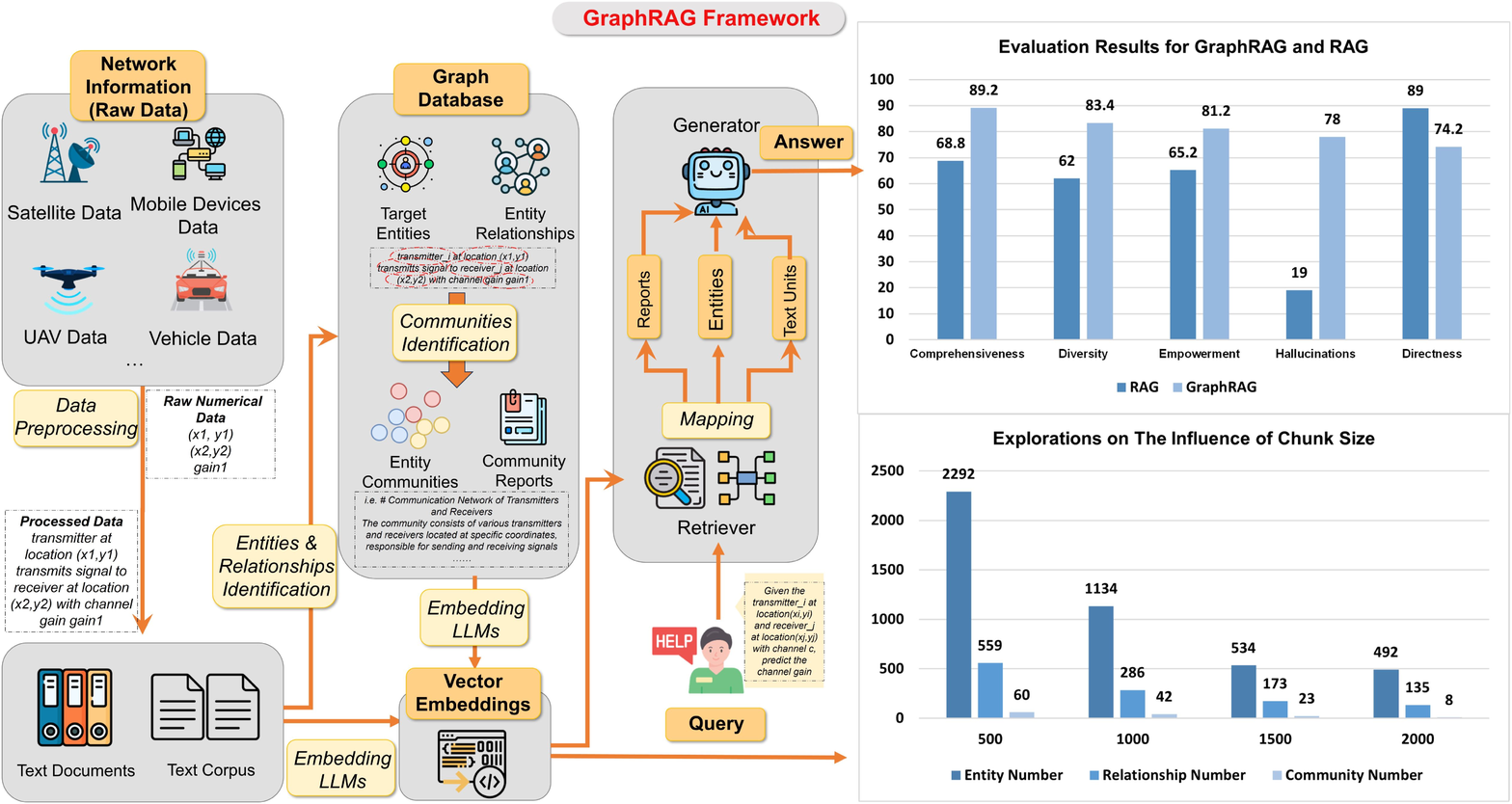}  
	\caption{In the GraphRAG framework, the most essential components include the database, the retriever, and the generator. The unique design of its integrated knowledge database and corresponding retrieval functions design equipped the GraphRAG framework with the ability to provide accurate and high-quality answers.} 
	\label{FIG:3}

\end{figure*}
\section{Proposed GraphRAG Framework and Construction Tutorial}
In this section, we investigate the impact of GraphRAG on network optimization, describing the key components of an LLM framework empowered by GraphRAG, and providing a step-by-step introduction to the integration of GraphRAG with networking.

\subsection{Framework Structure}
The design of our framework in Figure 3 is based on the key features of the GraphRAG system illustrated in Figure 2. The GraphRAG framework leverages the key strengths of GraphRAG and includes the following major components:

\begin{itemize}
\item\textbf{Knowledge Database:}
GraphRAG’s knowledge database is graph-structured, integrating unstructured, case-specific documents. Using LLMs such as GPT for entity and relationship extraction\footnote{{https://microsoft.github.io/graphrag/index/default\_dataflow/\#entity-relationship-extraction}}, we ensure that the graph captures network-relevant details. Additionally, data preprocessing is applied to handle pure numerical information (e.g., transmitter and receiver coordinates), creating vector embeddings for efficient retrieval.

\item\textbf{Retriever:}
GraphRAG's retriever, similar to that of the vanilla RAG framework, utilizes LLMs for language processing but is enhanced by a graph-structured database, enabling retrieval based on entities and their relationships. It also integrates data from community reports (e.g., related transmitters and receivers). After creating vector embeddings, the retriever extracts entities from queries, passing the retrieved information to the Generator for answer generation.

\item\textbf{Generator:}
The generator in GraphRAG shares a similar structure to that of the vanilla RAG framework. For consistency and accurate comparison in our case study, we use GPT-3.5-turbo with prompt engineering as the generator in both frameworks. 
\end{itemize}
\subsection{Tutorial: Constructing and Utilizing the GraphRAG Framework for Network Optimization}
This tutorial consists of two major parts. The first part introduces major steps of creating a knowledge base and knowledge graph. Then, the second part demonstrates how to leverage the GraphRAG framework for optimized information retrieval and user interaction in network environments.
\subsubsection{Knowledge Base Compilation and Knowledge Graph Construction}
 The setting up of GraphRAG involves compiling a comprehensive data set on the network, raw data preprocessing, and then structuring it into a knowledge graph for optimal retrieval and analysis. Follow these steps: 
\begin{itemize}
\item\textbf{Step 1: Data Collection and Preparation} Begin by gathering relevant documents and information on network-specific parameters, entities, and relationships. Typically, these documents include raw data from the network (i.e., channel information) or academic articles that contain field-specific information. This dataset will serve as the foundation for building a comprehensive, structured knowledge graph.

\item\textbf{Step 2: Entity Identification}
Convert key network elements (e.g., devices, users, services) into nodes. Annotate each node with relevant attributes, such as parameters (e.g., ``channel gain") or components (e.g., “receiver”). Incorporate diverse data sources to ensure complete representation

\item\textbf{Step 3: Relationship Mapping}
Define and map out relationships between entities to capture hierarchical and interactive dependencies within the network. For instance, represent a relationship such as “transmitter i transmits the signal to receiver j” to capture directional dependencies.

\item\textbf{Step 4: Graph Enrichment}
Leveraging algorithms such as the Hierarchical Leiden Algorithm\footnote{{https://neo4j.com/docs/graph-data-science/current/algorithms/leiden/}}, the system identifies and constructs communities that represent clusters of entities that are closely related (e.g., interconnected network devices) or interact frequently within the graph. After that, the system generates detailed reports for each community to summarize key information within the community. These reports provide overviews of the network’s structure from different perspectives.
\end{itemize}
\subsubsection{User Instruction and Information Retrieval}
Once the knowledge graph is built, you can begin retrieving and optimizing data based on user queries. The following steps demonstrate how to engage with GraphRAG effectively for optimized retrieval and data synthesis:
\begin{itemize}
\item\textbf{Step 1: User Query Interpretation}
The GraphRAG system, using the semantic extraction capabilities of LLMs, interprets user queries, identifying specific network optimization goals or issues. This ensures that each query is contextualized within the network's current state, drawing the most relevant data from the knowledge graph.

\item\textbf{Step 2: Graph-Based Information Retrieval}
Using GraphRAG’s graph structure, perform a semantic search\footnote{{https://microsoft.github.io/graphrag/query/overview/}} that goes beyond simple data retrieval to infer related information crucial for problem-solving. Advanced queries leverage entity relationships and community reports, allowing for detailed topic-based retrieval. For instance, when analyzing a network optimization problem, GraphRAG can categorize and summarize relevant concepts and parameters within each community, maximizing comprehension within the LLM's limited context window.

\item\textbf{Step 3: Interactive Refinement}
GraphRAG supports iterative user interactions, enabling users to refine their queries based on initial results. This step-by-step refinement process allows for more targeted information extraction and fine-tuned responses.
\end{itemize}

\section{Case Study: GraphRAG For Channel Gain Prediction}
In this section, we introduce the application of the proposed GraphRAG framework in network optimizations through a case study that utilizes GraphRAG to identify the channel gain given the CKM and the locations of the transmitter and receiver. The case study first demonstrates the effectiveness of the framework and then compares its performance against the baselines.

\begin{figure*}[!ht]       
	\centering
        \includegraphics[width=0.95\textwidth]{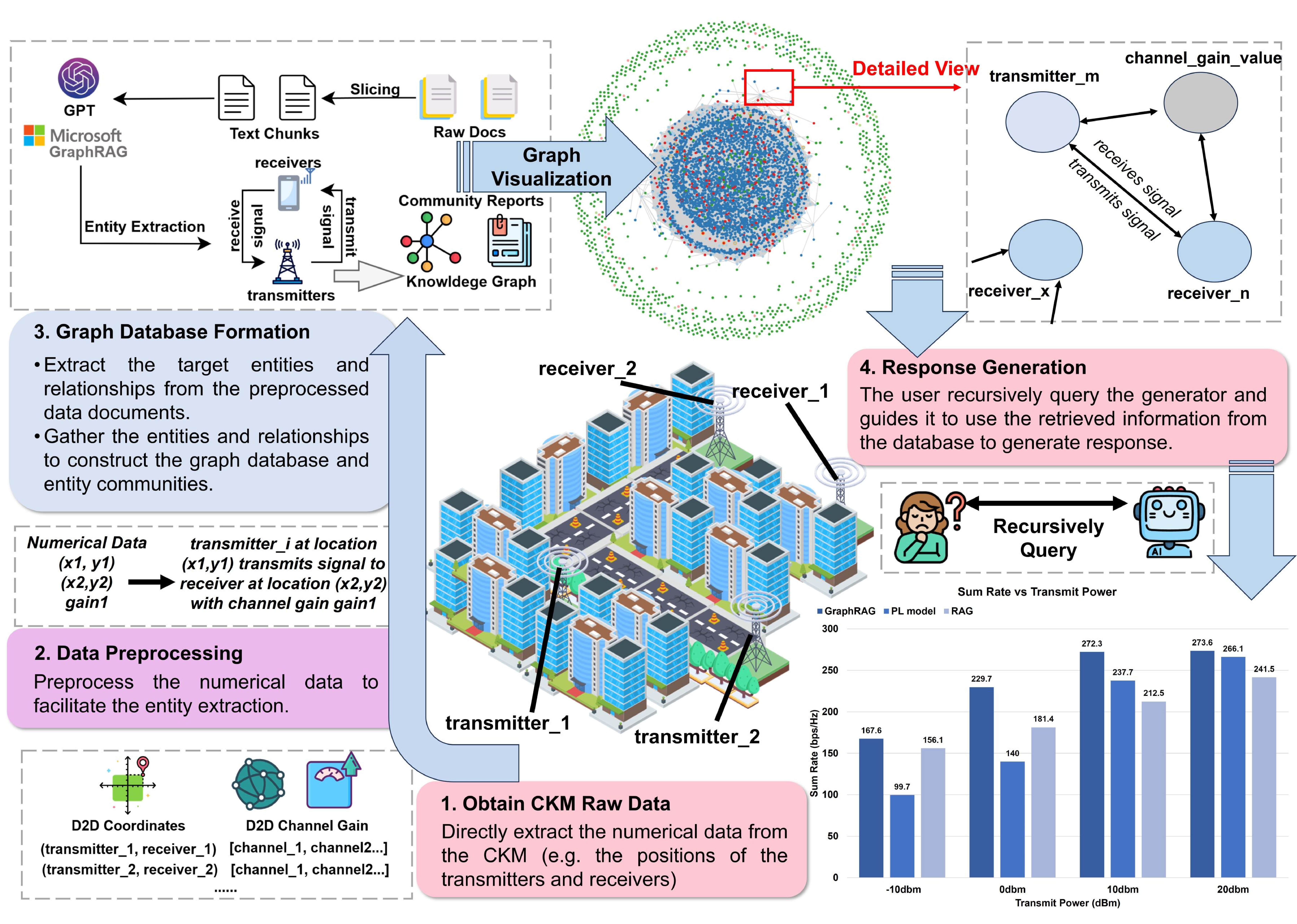}  
	\caption{Dataflow of the GraphRAG framework together with a visualization of the knowledge graph and a sum rate comparison between GraphRAG, Vanilla RAG, and PL model. The visualized knowledge graph is generated from the extracted entities and relationships. In the graph, the dots represent the entities, and the gray edges connecting them represent the relationships.}   
	\label{FIG:4}
\end{figure*}
\subsection{Motivation}
The rapid evolution of next-generation wireless networks introduces unprecedented challenges in terms of scalability, adaptability, and complexity in communication environments. As networks become more intricate, the existing modeling tools for spectrum maps have encountered several limitations \cite{motivation}. The growing complexity raised challenges toward clear representation of network topology and effective retrieval of network information, which demands a shift towards more intelligent and context-aware modeling techniques that can effectively integrate and utilize vast amounts of data. Traditional channel modeling methods, such as fitted path loss models, struggle to capture the nuanced dynamics of these systems. Traditional AI methods, while powerful, are often limited by their inability to reason over structured knowledge or dynamically adapt to rapidly changing network conditions. They primarily rely on predefined rules and lack the flexibility required to handle the multifaceted relationships inherent in modern wireless communication systems. 

To address these limitations, we propose a novel framework based on GraphRAG that leverages the reasoning capabilities of LLMs and the structural representation power of knowledge graphs. Compared to the traditional modeling approaches for spectrum maps, which often lack a clear and structured representation of data, GraphRAG's knowledge graph effectively maps the interconnections between devices within the spectrum map. This approach not only outperforms traditional AI methods in terms of contextual awareness and scalability but also offers a more structured, data-driven solution that can dynamically evolve with the network’s needs.
\subsection{Basic Validation}
In our experiments, we construct the vanilla RAG framework following the experiment settings in \cite{e}. The RAG framework is constructed using LangChain\footnote{{https://www.langchain.com/}} and OpenAI APIs\footnote{{https://platform.openai.com/docs/overview}}. For the GraphRAG framework, we utilize Microsoft's implementation of GraphRAG, bypassing LangChain to create the knowledge database. Both frameworks employ GPT-3.5 turbo as the underlying LLM for generating answers. We test token chunk sizes of 500, 1000, 1500, and 2000. As shown in Figure 3, bottom right, the number of extracted entities, relationships, and formulated communities increases as the text chunk size decreases from 2000 to 500 tokens. This indicates that smaller chunk sizes allow for more granular entity representations and a more intricate, interconnected knowledge graph. However, it is important to note that as chunk size decreases, the context available for entity and relationship extraction also reduces, potentially affecting the overall extraction quality, especially when the chunks are too small to capture relevant contextual information.

In addition to chunk size experimentation, we compared two frameworks by evaluating their responses to identical queries for optimization problem generation, using metrics from \cite{graphrag} and introducing ``hallucinations" to assess response faithfulness. Based on LLM (GPT-4) evaluations, this ``hallucinations" metric measures alignment with original sources, with higher scores indicating fewer hallucinations. The results averaged over ten trials show that GraphRAG outperforms vanilla RAG by ~30\% in comprehension, diversity, and empowerment, and achieves higher hallucination accuracy. This advantage stems from GraphRAG's interconnected knowledge graph, enabling richer context and nuanced responses. However, vanilla RAG performs better in directness, scoring 20\% higher due to its concise retrieval focus. For tasks requiring comprehensive analysis, GraphRAG demonstrates superior overall performance.

\subsection{Channel Gain Prediction Experiments}
Our experiment aims to explore the effectiveness of GraphRAG in solving network optimization problems. It builds upon the experiment that performs device-to-device (D2D) sub-band assignment using the CKM proposed in \cite{f}. Given the CKM containing the coordinates of the transmitters and receivers, as well as the corresponding channel gains, we construct the knowledge database for the GraphRAG system using the ray-tracing D2D data from the CKM. We then recursively query the generator, guiding it to infer the channel gain based on the coordinates of the transmitter and receiver using information retrieved from the knowledge database. 
\begin{itemize}

\item{\textbf{Knowledge Database Construction and Experiment Settings:}}
The construction of the knowledge database begins with formatting the raw ray-tracing D2D data. We read the locations and channel gain data for each transmitter-receiver pair in the raw data. Each time we encounter a transmitter (or receiver) at a new location, we assign an identical label i or j to it (e.g. when we meet a transmitter at a new location and the last labeled transmitter is transmitter\_100, then we label this transmitter as transmitter\_101). Afterward, we store the labeled transmitter-receiver pairs along with their channel gain information in a text document. The purpose of the formatting process is to create clearer entity representations from raw digital data while reducing duplicated entities. 
With the channel knowledge data formatted, we forward it to the LLM to perform entities and relationship extraction. We use GPT-3.5-turbo with a chunk size of 1000 for extraction. Since the total number of entities and relationships in the knowledge graph of the entire CKM would be too large and the cost of fully extracting such a knowledge map would be too high, we randomly select 1/10 of the data in the CKM to perform entities and relationship extraction. We set the entity types to ``transmitter", ``receiver", ``channel gain", ``coordinate", and ``value" to guide the entity extraction process. After extraction, we obtained 25,625 entities and 64,527 relationships. We use the visualization tool\footnote{{https://noworneverev.github.io/graphrag-visualizer/}} to provide a visualization of the knowledge graph in Figure 4. 

We conduct the channel gain prediction by replacing the Deep Neural Network (DNN) in \cite{f} with the GraphRAG framework to predict the channel gain given the locations of the transmitters and corresponding receivers. For both the GraphRAG framework and vanilla RAG framework, the predictions are made using the GPT-3.5-turbo generator. We design prompts to guide the LLM generator in leveraging the retrieved information to produce formatted responses, allowing us to extract the channel gain predicted along with corresponding transmitter-receiver pairs.

\item{\textbf{Performance Evaluation:}}
Next, in this case study, GraphRAG's performance is demonstrated through the achievable sum rate. For each transmitter-receiver pair, we extract the highest predicted channel gain from the generator’s output and format it for sum rate calculation. As shown in Figure 4, GraphRAG consistently outperforms both the PL model and vanilla RAG across different transmit power levels. At lower transmit power, both GraphRAG and vanilla RAG show an advantage over the PL model. However, as the transmit power increases to 20 dBm, vanilla RAG's performance plateaus, whereas GraphRAG maintains a significant edge. This superior performance of GraphRAG can be attributed to its graph-structured knowledge database, which allows it to better capture relationships between network entities and leverage contextual information more effectively. This detailed understanding enables GraphRAG to infer more accurate channel gains. These results are consistent with our earlier findings in Figure 4, where GraphRAG's enhanced entity and relationship extraction contributes to its overall performance in complex network environments.

\end{itemize}

\section{Future Works}
In this section, we outline three key directions for advancing the application of knowledge-graphs-empowered RAG in networking.

\textbf{Robust Graph Updates:} A major limitation of current RAG frameworks with knowledge graphs is the lack of a robust and efficient update mechanism, essential for maintaining accuracy and relevance in dynamic networking environments. Developing a robust and cost-effective update system is critical for real-time adaptability, data integrity, and broader applicability in complex scenarios.

\textbf{Addressing Hallucination Issues:} While knowledge-graphs-empowered RAG performs better than standard RAG in reducing hallucinations, it still faces challenges in this area. Enhancing methods to further minimize hallucination could improve its reliability for networking tasks that demand high precision.

\textbf{Information Security:} Building knowledge-graphs-empowered RAG’s database requires regular interaction with LLM agents, posing potential data leakage risks. Future research should prioritize robust security measures to safeguard the data.

\section{Conclusion}
In this article, we have discussed RAG's limitations and explained how knowledge-graphs-empowered RAG advances upon vanilla RAG, both structurally and in terms of the benefits it offers. We have then proposed the GraphRAG framework for networking, accompanied by a step-by-step guide on its construction. Finally, we have assessed the effectiveness of our framework through a case study and suggested potential future research directions for applying knowledge-graphs-empowered RAG in networking.

\bibliographystyle{IEEEtran}
\bibliography{mylib}

\begin{thebibliography}{10}
\providecommand{\url}[1]{#1}
\csname url@samestyle\endcsname
\providecommand{\newblock}{\relax}
\providecommand{\bibinfo}[2]{#2}
\providecommand{\BIBentrySTDinterwordspacing}{\spaceskip=0pt\relax}
\providecommand{\BIBentryALTinterwordstretchfactor}{4}
\providecommand{\BIBentryALTinterwordspacing}{\spaceskip=\fontdimen2\font plus
\BIBentryALTinterwordstretchfactor\fontdimen3\font minus \fontdimen4\font\relax}
\providecommand{\BIBforeignlanguage}[2]{{%
\expandafter\ifx\csname l@#1\endcsname\relax
\typeout{** WARNING: IEEEtran.bst: No hyphenation pattern has been}%
\typeout{** loaded for the language `#1'. Using the pattern for}%
\typeout{** the default language instead.}%
\else
\language=\csname l@#1\endcsname
\fi
#2}}
\providecommand{\BIBdecl}{\relax}
\BIBdecl

\bibitem{demllm}
R.~Zhang, J.~He, X.~Luo, D.~Niyato, J.~Kang, Z.~Xiong, Y.~Li, and B.~Sikdar, ``Toward democratized generative ai in next-generation mobile edge networks,'' 2024.

\bibitem{idnllm}
E.-D. Jeong, H.-G. Kim, S.~Nam, J.-H. Yoo, and J.~W.-K. Hong, ``S-witch: Switch configuration assistant with llm and prompt engineering,'' in \emph{NOMS 2024-2024 IEEE Network Operations and Management Symposium}, 2024, pp. 1--7.

\bibitem{rag}
P.~Lewis, E.~Perez, A.~Piktus, F.~Petroni, V.~Karpukhin, N.~Goyal, H.~K\"{u}ttler, M.~Lewis, W.-t. Yih, T.~Rockt\"{a}schel, S.~Riedel, and D.~Kiela, ``Retrieval-augmented generation for knowledge-intensive nlp tasks,'' in \emph{Proceedings of the 34th International Conference on Neural Information Processing Systems}, ser. NIPS '20.\hskip 1em plus 0.5em minus 0.4em\relax Red Hook, NY, USA: Curran Associates Inc., 2020.

\bibitem{c}
G.~Sun, W.~Xie, D.~Niyato, H.~Du, J.~Kang, J.~Wu, S.~Sun, and P.~Zhang, ``Generative ai for advanced uav networking,'' \emph{IEEE Network}, pp. 1--1, 2024.

\bibitem{6g}
W.~Saad, M.~Bennis, and M.~Chen, ``A vision of 6g wireless systems: Applications, trends, technologies, and open research problems,'' \emph{IEEE Network}, vol.~34, no.~3, pp. 134--142, 2020.

\bibitem{satelite}
R.~Zhang, H.~Du, Y.~Liu, D.~Niyato, J.~Kang, Z.~Xiong, A.~Jamalipour, and D.~In~Kim, ``Generative ai agents with large language model for satellite networks via a mixture of experts transmission,'' \emph{IEEE Journal on Selected Areas in Communications}, vol.~42, no.~12, pp. 3581--3596, 2024.

\bibitem{lowcarbon}
J.~Wen, R.~Zhang, D.~Niyato, J.~Kang, H.~Du, Y.~Zhang, and Z.~Han, ``Generative ai for low-carbon artificial intelligence of things with large language models,'' 2024.

\bibitem{mobileai}
Z.~Chen, Q.~Sun, N.~Li, X.~Li, Y.~Wang, and C.-L. I, ``Enabling mobile ai agent in 6g era: Architecture and key technologies,'' \emph{IEEE Network}, vol.~38, no.~5, pp. 66--75, 2024.

\bibitem{ragcontext}
N.~F. Liu, K.~Lin, J.~Hewitt, A.~Paranjape, M.~Bevilacqua, F.~Petroni, and P.~Liang, ``Lost in the middle: How language models use long contexts,'' \emph{Transactions of the Association for Computational Linguistics}, vol.~12, pp. 157--173, 2024.

\bibitem{ragacc}
F.~Shi, X.~Chen, K.~Misra, N.~Scales, D.~Dohan, E.~Chi, N.~Sch\"{a}rli, and D.~Zhou, ``Large language models can be easily distracted by irrelevant context,'' in \emph{Proceedings of the 40th International Conference on Machine Learning}, ser. ICML'23.\hskip 1em plus 0.5em minus 0.4em\relax JMLR.org, 2023.

\bibitem{ragsummary}
T.~Goodwin, M.~Savery, and D.~Demner-Fushman, ``Flight of the {PEGASUS}? comparing transformers on few-shot and zero-shot multi-document abstractive summarization,'' in \emph{Proceedings of the 28th International Conference on Computational Linguistics}, D.~Scott, N.~Bel, and C.~Zong, Eds.\hskip 1em plus 0.5em minus 0.4em\relax Barcelona, Spain (Online): International Committee on Computational Linguistics, Dec. 2020, pp. 5640--5646.

\bibitem{graphrag}
D.~Edge, H.~Trinh, N.~Cheng, J.~Bradley, A.~Chao, A.~Mody, S.~Truitt, and J.~Larson, ``From local to global: A graph rag approach to query-focused summarization,'' 2024.

\bibitem{motivation}
H.~A.~H. Alobaidy, M.~Jit~Singh, M.~Behjati, R.~Nordin, and N.~F. Abdullah, ``Wireless transmissions, propagation and channel modelling for iot technologies: Applications and challenges,'' \emph{IEEE Access}, vol.~10, pp. 24\,095--24\,131, 2022.

\bibitem{e}
R.~Zhang, H.~Du, Y.~Liu, D.~Niyato, J.~Kang, S.~Sun, X.~Shen, and H.~V. Poor, ``Interactive ai with retrieval-augmented generation for next generation networking,'' \emph{IEEE Network}, vol.~38, no.~6, pp. 414--424, 2024.

\bibitem{f}
Y.~Zeng and X.~Xu, ``Toward environment-aware 6g communications via channel knowledge map,'' \emph{IEEE Wireless Communications}, vol.~28, no.~3, pp. 84--91, 2021.

\end{thebibliography}

\end{CJK}
\end{document}